\documentclass[preprint,showpacs,preprintnumbers,amsmath,amssymb,letterpaper,prd]{revtex4}

\begin{document}
\sloppy

\title{Mixed colour states in QCD confining vacuum}
\author{V. I. Kuvshinov}%
\email{v.kuvshinov@sosny.bas-net.by}
\affiliation{JIPNR, National Academy of Science, 220109 Belarus, Minsk, Acad. Krasin str. 99}
\author{P. V. Buividovich}
\email{buividovich@tut.by}
\affiliation{Belarusian State University, 220080 Belarus, Minsk, Nezalezhnasti av. 4}
\date{4 June 2006}
\begin{abstract}
We show that confinement of spinless heavy quarks in fundamental representation of $SU(N_{c})$ gauge group can be treated as decoherence of pure colour state into a white mixture of states. Decoherence rate is found to be proportional to the tension of QCD string and the distance between colour charges. The purity of colour states is calculated.
\end{abstract}
\maketitle

 One of the most popular test objects used to detect quark confinement is a Wilson loop calculated in some representation of the gauge group. Measurements of the Wilson loops allow one to investigate the interaction potential between static quarks in different representations of the gauge group. For instance, Wilson area law corresponds to QCD string with constant tension \cite{Wilson:74}. Most numerical simulations indicate that Wilson loops indeed obey the Wilson area law at intermediate distances $\sim 0.2 \ldots 1$ fm \cite{Bali:00, Deldar:00}.

 Measurements of Wilson loops allow one to obtain, for example, string tensions or field configurations around static colour charges \cite{Steffen:03, Dosch:02}. In this paper we will investigate what can be inferred about the behavior of the colour states of quarks themselves, provided Wilson loops decay. Usually white wave functions of hadrons are constructed as gauge-invariant superpositions of quark colour states. This paper is an attempt  to treat the confinement of quarks from a quantum-optical point of view and to demonstrate that white objects can be also obtained as white mixtures of states described by the density matrix. In this case QCD confining vacuum can be considered as the environment in the language of quantum optics.

 Let us clarify this analogy. Suppose that we have some quantum system which interacts with the environment. The  system itself in general can not be described by the state vector \cite{PeresQuantumTheory, HaakeQuantumChaos}, as only the joint state vector of the system and the environment can be considered. But the system itself can be described by the density matrix. Such states of the system are called the mixed states \cite{PeresQuantumTheory, HaakeQuantumChaos}. The density matrix of the system can be obtained by calculating the density matrix of the system and the environment together and after that taking traces with respect to those degrees of freedom which belong to the environment \cite{PeresQuantumTheory, HaakeQuantumChaos}.

 One uses essentially the same method when calculating the Wilson loops in QCD, either lattice or continuum: colour states of heavy quarks are averaged over the states of gluons and sea quarks. As a result no individual colour states are considered, but rather colour states of particles connected by QCD string. In this paper we will apply the quantum-mechanical procedure mentioned above to the colour states of heavy quarks: consider the density matrix of colour charge and average it over the QCD confining vacuum. Finally we will demonstrate that the purity \cite{PeresQuantumTheory} of the density matrix corresponding to colour degrees of freedom can be expressed in terms of Wilson loops.

 Interactions with the environment result in decoherence and relaxation of quantum superpositions \cite{PeresQuantumTheory, HaakeQuantumChaos}, so that the information on the initial state of the quantum system is lost after sufficiently large time. Here the analogy between QCD vacuum and the environment can be continued: information on colour states is also lost in QCD vacuum due to confinement phenomenon. 

 In order to demonstrate the emergence of white mixed states let us investigate the following example. Consider propagation of heavy spinless quark along some closed path $\gamma$ and suppose that in the point $x_{in}$ of the loop $\gamma$ the colour state of the quark is described by the state vector $| \phi_{in} \rangle$ (thus it is a pure state).  The amplitude of such process in Feynman-Schwinger representation and in quenched approximation factorizes into the kinetic part and the parallel transport operator \cite{Steffen:03, Dosch:02}, hence in semiclassical approximation the colour state vector of the quark is parallel transported along the loop $\gamma$:
\begin{eqnarray}
\label{ParallelTransport}
|\phi (\gamma) \rangle = \mathcal{P} \exp \left( i \int \limits_{\gamma} dx^{\mu} \hat{A}_{\mu} \right) | \phi_{in} \rangle
\end{eqnarray}
where by kets we denote colour state vectors (unit vectors in $N_{c}$-dimensional complex space), $\mathcal{P}$ is the path-ordering operator and $\hat{A}_{\mu}$ is the gauge field vector. In order to consider mixed states we use conventional quantum-mechanical definition of the density matrix, treating QCD vacuum as the environment: 
\begin{eqnarray}
\label{DMDefinition}
\hat{\rho}(\gamma) = \langle \langle \: | \phi(\gamma) \rangle \langle \phi(\gamma) | \: \rangle \rangle
\end{eqnarray}
 where by $ \langle \langle \ldots \rangle \rangle$ we denote the vacuum expectation value.

 According to the equation (\ref{ParallelTransport}) the colour density matrix $\hat{\rho} (\gamma)$ is:
\begin{eqnarray}
\label{CDM1}
 \hat{\rho} (\gamma) =  \langle \langle \: \mathcal{P} \exp \left( i \int \limits_{\gamma} dx^{\mu} \hat{A}_{\mu} \right)
 | \phi_{in} \rangle \langle \phi_{in} | \nonumber \\
 \mathcal{P} \exp \left( i \int \limits_{\gamma} dx^{\mu} \hat{A}_{\mu} \right)^{\dag}
    \: \rangle \rangle
\end{eqnarray}

The colour density matrix is hermitian and obeys the constraint ${\rm Tr} \: \hat{\rho} = 1$, therefore we can decompose it into the pieces which transform under adjoint and trivial representations of the gauge group as $\hat{\rho} =  N^{-1}_{c} \hat{I} +  \hat{T}_{a} \rho^{a}$, where $\hat{T}_{a}$, $a = 1 \ldots N^{2}_{c} - 1$ are the generators of $SU(N_{c})$ Lie algebra normalized as ${\rm Tr} \: \left(\hat{T}_{a} \hat{T}_{b} \right) = \delta_{a b}$. Making use of this decomposition we can rewrite the expression (\ref{CDM1}) in the following form:
\begin{eqnarray}
\label{DensityMatrixFinal}
    \hat{\rho} \left( \gamma \right) = \langle \langle \: N^{-1}_{c} \hat{I} + \nonumber \\ +  \hat{T}_{a}
    \langle \phi_{in} |\hat{T}^{b} | \phi_{in} \rangle
     \mathcal{P} \exp \left( \int \limits_{\gamma} dx^{\mu} \hat{A}^{adj}_{\mu} \right) {}^{a}_{b} \:  \rangle \rangle
\end{eqnarray}
where the path-ordered exponent is calculated in the adjoint representation of the gauge group. As we average the density matrix only over gluonic degrees of freedom and the only term in (\ref{DensityMatrixFinal}) which contains gauge field is the path-ordered exponent, the expression for the density matrix (\ref{DensityMatrixFinal}) becomes:
\begin{eqnarray}
\label{DensityMatrixFinal1}
    \hat{\rho} \left( \gamma \right) =  N^{-1}_{c} \hat{I} + \nonumber \\ +
     \hat{T}_{a}
    \langle \phi_{in} |\hat{T}^{b} | \phi_{in} \rangle
    \langle \langle \: \mathcal{P} \exp \left( \int \limits_{\gamma} dx^{\mu} \hat{A}^{adj}_{\mu} \right) {}^{a}_{b}  \: \rangle \rangle
\end{eqnarray}

 Expectation value of path-ordered exponent in (\ref{DensityMatrixFinal1}) is proportional to the identity due to gauge invariance (we assume no gauge fixing here, which is valid for lattice simulations), therefore we can write it as follows:
\begin{eqnarray}
\label{LoopViaTrace}
   \langle \langle \: \mathcal{P} \exp \left( \int \limits_{\gamma} dx^{\mu} \hat{A}^{adj}_{\mu} \right) {}^{a}_{b} \: \rangle \rangle = \nonumber \\ =
\delta^{a}_{b} \left({N^{2}_{c} - 1} \right)^{-1} \langle \langle \: {\rm Tr} \: \mathcal{P} \exp \left( \int \limits_{\gamma} dx^{\mu} \hat{A}^{adj}_{\mu} \right)  \: \rangle \rangle = \nonumber \\ = \delta^{a}_{b} W_{adj} \left( \gamma \right)
\end{eqnarray} 
where $W_{adj} \left( \gamma \right)$ is the Wilson loop in the adjoint representation.

Finally after taking into account (\ref{LoopViaTrace}) we obtain for the colour density matrix of the colour charge which was parallel transported along the loop $\gamma$:
\begin{eqnarray}
\label{ChargeDensityMatrixAfterAveraging}
    \hat{\rho} \left(\gamma \right) =  N^{-1}_{c} \hat{I} + \left(| \phi_{in} \rangle \langle \phi_{in} | -  N^{-1}_{c} \hat{I} \right) W_{adj} \left( \gamma \right)
\end{eqnarray}
It is important to note that this colour density matrix transforms covariantly under gauge transformations, i.e.
$\hat{\rho} \left(\gamma \right) \rightarrow \hat{U}(x_{in}) \hat{\rho} \left(\gamma \right) \hat{U}^{\dag}(x_{in})$, because vacuum expectation value of a gauge-variant quantity is a gauge-invariant quantity, if one assumes no gauge fixing.

 Expression (\ref{ChargeDensityMatrixAfterAveraging}) shows that if the Wilson loop in the adjoint representation decays, the colour density matrix obtained as a result of parallel transport along the loop $\gamma$ tends to white colourless mixture with $\hat{\rho} = N^{-1}_{c} \hat{I}$, where all colour states are mixed with equal probabilities and all information on the initial colour state is lost. But Wilson loop decay points at confinement of colour charges, therefore the stronger are the colour charges confined, the quicker their pure colour states transform into white mixtures. It is important that the path $\gamma$ is closed, which means that actually one observes particle and antiparticle rather than a single colour charge.

 As the Wilson area law typically holds for the Wilson loop \cite{Wilson:74}, we can obtain an explicit expression for the density matrix. Here it is convenient to choose the rectangular loop $\gamma_{R \times T}$ which stretches time $T$ and distance $R$:
\begin{eqnarray}
\label{DensityMatrixArea}
 \hat{\rho} \left(\gamma_{R \times T} \right) = N^{-1}_{c} \hat{I} + \nonumber \\ + \left(| \phi_{in} \rangle \langle \phi_{in} | -  N^{-1}_{c} \hat{I} \right)
 \exp \left( - \sigma_{adj} R T \right)
\end{eqnarray}
where $\sigma_{adj}$ is the string tension between charges in the adjoint representation.

 Finally we can obtain the decoherence rate, which is introduced using the concept of purity $p = {\rm Tr} \: \hat{\rho}^{2}$ \cite{PeresQuantumTheory, HaakeQuantumChaos}. For pure states the purity is equal to one, for mixed states the purity is always less than one. Purity is a gauge-invariant quantity, which follows immediately from the transformation law of the colour density matrix (\ref{ChargeDensityMatrixAfterAveraging}).  For our colour density matrix (\ref{DensityMatrixArea}) the purity is:
\begin{eqnarray}
\label{PurityInStochasticVacuum}
 p\left(T, R \right) = N^{-1}_{c} + \left(1 - N^{-1}_{c} \right) \exp \left( - 2 \sigma_{adj} R T \right)
\end{eqnarray}
As $T$ or $R$ tend to zero, the purity tends to one, which corresponds to the pure state with the density matrix $\hat{\rho} = | \phi_{in} \rangle \langle \phi_{in} |$. As the product $R T$ tends to infinity, the purity tends to $N_{c}^{-1}$, which corresponds to white mixed state with the density matrix $\hat{\rho} = N_{c}^{-1} \hat{I}$. Purity decay rate $T^{-1}_{dec} = 2 \sigma_{adj} R $, where $T_{dec}$ is the characteristic decoherence time, is proportional to the string tension and the distance $R$. It can be inferred from this expression that the stronger is the quark-antiquark pair coupled by QCD string or the larger is the distance between quark and antiquark, the quicker information about colour states is lost as a result of interactions with QCD vacuum.

 Thus we have considered the evolution of a single colour charge interacting with QCD vacuum. As it is suggested by the expression (\ref{PurityInStochasticVacuum}), confinement indeed can be regarded as a result of decoherence processes. Decoherence rate is twice the tension of QCD string. It could be interesting to extend the arguments presented above to the case of dynamical quarks.

%\bibliography{MyBibliography}

\begin{thebibliography}{7}
\expandafter\ifx\csname natexlab\endcsname\relax\def\natexlab#1{#1}\fi
\expandafter\ifx\csname bibnamefont\endcsname\relax
  \def\bibnamefont#1{#1}\fi
\expandafter\ifx\csname bibfnamefont\endcsname\relax
  \def\bibfnamefont#1{#1}\fi
\expandafter\ifx\csname citenamefont\endcsname\relax
  \def\citenamefont#1{#1}\fi
\expandafter\ifx\csname url\endcsname\relax
  \def\url#1{\texttt{#1}}\fi
\expandafter\ifx\csname urlprefix\endcsname\relax\def\urlprefix{URL }\fi
\providecommand{\bibinfo}[2]{#2}
\providecommand{\eprint}[2][]{\url{#2}}

\bibitem[{\citenamefont{Wilson}(1974)}]{Wilson:74}
\bibinfo{author}{\bibfnamefont{K.~G.} \bibnamefont{Wilson}},
  \bibinfo{journal}{Physical Review D} \textbf{\bibinfo{volume}{10}},
  \bibinfo{pages}{2445} (\bibinfo{year}{1974}).

\bibitem[{\citenamefont{Bali}(2000)}]{Bali:00}
\bibinfo{author}{\bibfnamefont{G.~S.} \bibnamefont{Bali}},
  \bibinfo{journal}{Physical Review D} \textbf{\bibinfo{volume}{62}},
  \bibinfo{pages}{114503} (\bibinfo{year}{2000}).

\bibitem[{\citenamefont{Deldar}(2000)}]{Deldar:00}
\bibinfo{author}{\bibfnamefont{S.}~\bibnamefont{Deldar}},
  \bibinfo{journal}{Physical Review D} \textbf{\bibinfo{volume}{62}},
  \bibinfo{pages}{034509} (\bibinfo{year}{2000}).

\bibitem[{\citenamefont{Shoshi et~al.}(2003)\citenamefont{Shoshi, Steffen,
  Dosch, and Pirner}}]{Steffen:03}
\bibinfo{author}{\bibfnamefont{A.~I.} \bibnamefont{Shoshi}},
  \bibinfo{author}{\bibfnamefont{F.~D.} \bibnamefont{Steffen}},
  \bibinfo{author}{\bibfnamefont{H.~G.} \bibnamefont{Dosch}}, \bibnamefont{and}
  \bibinfo{author}{\bibfnamefont{H.~J.} \bibnamefont{Pirner}},
  \bibinfo{journal}{Physical Review D} \textbf{\bibinfo{volume}{68}},
  \bibinfo{pages}{074004} (\bibinfo{year}{2003}).

\bibitem[{\citenamefont{Giacomo et~al.}(2002)\citenamefont{Giacomo, Dosch,
  Shevchenko, and Simonov}}]{Dosch:02}
\bibinfo{author}{\bibfnamefont{A.~D.} \bibnamefont{Giacomo}},
  \bibinfo{author}{\bibfnamefont{H.~G.} \bibnamefont{Dosch}},
  \bibinfo{author}{\bibfnamefont{V.~I.} \bibnamefont{Shevchenko}},
  \bibnamefont{and} \bibinfo{author}{\bibfnamefont{Y.~A.}
  \bibnamefont{Simonov}}, \bibinfo{journal}{Physics Reports}
  \textbf{\bibinfo{volume}{372}}, \bibinfo{pages}{319} (\bibinfo{year}{2002}).

\bibitem[{\citenamefont{Peres}(1995)}]{PeresQuantumTheory}
\bibinfo{author}{\bibfnamefont{A.}~\bibnamefont{Peres}},
  \emph{\bibinfo{title}{Quantum Theory: Concepts and Methods}}
  (\bibinfo{publisher}{Kluwer}, \bibinfo{address}{Dordrecht},
  \bibinfo{year}{1995}).

\bibitem[{\citenamefont{Haake}(1991)}]{HaakeQuantumChaos}
\bibinfo{author}{\bibfnamefont{F.}~\bibnamefont{Haake}},
  \emph{\bibinfo{title}{Quantum signatures of chaos}}
  (\bibinfo{publisher}{Springer-Verlag}, \bibinfo{address}{Berlin},
  \bibinfo{year}{1991}).
\end{thebibliography}
%\bibliographystyle{apsrev}

\end{document}